\documentstyle[12pt]{article}
\normalbaselines
 
\begin{document}
 \bibliographystyle{unsrt}
 \vbox{\vspace{6mm}}

\begin{center}
 {{\bf \Large f-OSCILLATORS AND NONLINEAR COHERENT STATES}} \\[7mm]

V. I. Man'ko,$^{1,2}$ G. Marmo,$^{3,4}$ E. C. G. Sudarshan,$^{5}$  
and F. Zaccaria$^{3,4}$ \\[2mm]

{\em $^{1}$Osservatorio Astronomico di Capodimonte\\
 Via Moiariello 16, I-80131 Napoli, Italy}\\[5mm]
{\em $^{2}$Lebedev Physical Institute\\
 Leninsky Pr., 53, Moscow 117924, Russia}\\[5mm]
{\em $^{3}$Dipartimento di Scienze Fisiche, Universit\`{a} di Napoli\\
$^{4}$Istituto Nazionale di Fisica Nucleare, Sezione di Napoli\\
Mostra d'Oltremare, Pad. 19 I-80125 Napoli, Italy}\\[5mm]
{\em $^{5}$Physics Department, Center for Particle Physics\\
University of Texas, Austin, Texas, 78712 USA}  
\end{center}

\begin{abstract}
The notion of f--oscillators generalizing q--oscillators is introduced.
For classical and quantum cases, an interpretation of the
f--oscillator is provided  as corresponding to a special nonlinearity
of vibration for which the frequency of oscillation depends on the 
energy. The f--coherent states (nonlinear coherent states) 
generalizing q--coherent states are constructed. Applied to quantum 
optics, photon distribution function, photon number means, and dispersions 
are calculated for the f--coherent states as well as the Wigner function 
and Q--function. As an example, it is shown how this nonlinearity may 
affect the Planck distribution formula.

\end{abstract}

\section{Introduction}

\indent\indent
In quantum physics, harmonic oscillators are synonymous with creation 
and annihilation operators. For this reason, in the first attempts to 
realize Hopf algebras (quantum groups) in terms of creation and 
annihilation operators (a generalization of the Jordan--Schwinger map)
the resulting oscillators were named q--oscillators. This pervading 
property of the oscillator formalism in many physical situations 
has induced a lot of interest in looking for physical consequences,
where honest oscillators are replaced by q-deformed ones (partition
functions, field theories, nonlinear optics, etc.).

Coherent states, defined through creation and annihilation operators,
provide us with a beautiful connection between quantum and classical
oscillators. The notion of coherent states~[1--3]  
%\cite{gla},\cite{sud1},\cite{kla1}
permitted the use of language  and intuition developed from the
study of the classical mechanics of  harmonic oscillators in order to
treat their quantum counterpart, because the trajectory of the center
of quantum coherent packet is the same as the classical trajectory
and the width of the packet is the minimal possible one in the frame 
of Heisenberg uncertainty relation~\cite{heis}. The notion of coherent 
state turned out to be appropriate also to describe simple quantum
systems like spin~\cite{rad} and cyclotron motion of a charge in 
magnetic field~\cite{malman}. On the other hand, the notion of the  
quantum q--oscillator~\cite{bie,mc} was 
interpreted~\cite{sol1,sol2} as a nonlinear oscillator
with a very specific type of nonlinearity, in which the frequency
of vibration depends on the energy of these vibrations through the
hyperbolic cosine  function containing a parameter of nonlinearity.

This interpretation of q--oscillators becomes obvious if one used the
classical  counterpart of the original quantum q--oscillators. This
observation suggests that  there might exist other types of nonlinearity
for which the frequency of oscillation varies with the amplitude in a 
manner different from the $~cosh $--dependence; we will label this 
dependence by a function $~f$. Such classical oscillators (and their 
quantum partners) may be called f--oscillators~\cite{sol1}.
It is interesting to consider the statistical mechanics of a gas
of deformed oscillators (free energy, partition function behaviour) 
and to compare it with the one associated with standard oscillators.

The  problems have obvious counterparts in quantum mechanics. Here, 
the role of the phase diagram is played by the eigenstates of the
Hamiltonian. For stationary systems, one could consider such changes 
of the Hamiltonian, which is an integral of motion, that produce new 
Hamiltonian which is some function of the initial one. Then if there 
are no degeneracies in the spectra for the initial Hamiltonian, the 
eigenstates of the new Hamiltonian coincide with the old ones. But 
for the new system the energy spectra are different. This produces 
time evolution of the phase factors of the eigenstates such that these 
vary with different velocities in complete analogy to the classical 
motion of the corresponding deformed classical systems, moving along 
their trajectories in phase space with reparametrized velocities.

In fact, in more general situations the new quantum systems having the
same stationary eigenstates as the initial ones possess the new Hamiltonian
which is a function of the usually commuting time independent integrals of
motion (complete set of observables). This is the analogue of the new
classical Hamiltonian which was deformed using reparametrizations 
depending on all available classical invariants which are in involution.

The aim of our paper is to study the behaviour of  classical and
quantum systems belonging to the subclass described above, and to 
clarify the role of nonlinearities corresponding to these systems. This 
goal is motivated by the fact that q--oscillator belongs to the system of 
the subclass corresponding to the specific q--nonlinearity~\cite{sol1}.

In classical case, we consider simple linear systems (oscillators) and
their deformations producing nonlinear integrable systems as well as
symmetries based on the nonlinear noncanonical transform of the
conjugate variables preserving the vectorfield. In the quantum
counterpart, we study such systems which differ in Hamiltonian but
have the same set of eigenstates. In these cases, we analyze the
possibility of extending the notion of coherent states of usual
harmonic oscillator to the case of  f--oscillators. Algebraic 
extensions of the notion of q--oscillator coherent states have been
discussed in~\cite{dask,quesne} and applications in~\cite{mukunda}.
The particular case of f--coherent states called also as nonlinear
coherent states for the function $~f~$ expressed in terms of
Laguerre polynomials was shown to be created for trapped ion 
in~\cite{wvogel}. Shortly f--oscillators were discussed in~\cite{wigsym}.

We will study some physical consequences of the existence of f--coherent
states like the change of the particle distribution function, the
possibility of having super- or sub-Poissonian statistics, influence
of f--nonlinearity in the black body radiation formula with the
particular example of the q--oscillators.
As a particular example we will apply these results to q--nonlinear
systems (q--oscillators) and will show that for q--coherent states 
there exists sub-Poissonian statistics which mean that q--nonlinearity 
of their fields decreases the fluctuation of the particle number in a
q--coherent state.

The next three sections (2, 3 and 4) illustrate in detail the situation 
in classical mechanics with the examples of one-dimensional,
two-dimensional and three-dimensional 
oscillators. The introduction of the
one-dimensional quantum analogues (the quantum f--oscillator) is given 
in Section~5 and in the following one (Section~6) some algebraic 
relations are shown for the operators describing them. Then the 
eigenstates of a one-mode f--annihilation operator 
(f--coherent states) are considered in the Fock space (Section~7) 
as well as in different representations (Wigner and Husimi) in Section~8. 
Properties of such states are studied in Sections~9, 10 and 11, namely, 
their evolution and completeness relations with a remark on the Stone--von 
Neumann theorem, with the main end of underlining that they are not 
always coherent in the ordinary sense~\cite{kla2}. Extensions to many modes
are shown to be possible not only when they are independent of each
other but also when there is a nonlinear coupling among them (Section~12).
The current setup of quantum mechanics  having been maintained, an 
application of the above objects in quantum optics is given in the 
last Sections~13, 14 and 15 obtaining new photon distributions, squeezing 
in the nonlinear coherent states, correlation of quadratures, and Planck 
formula deformation.

\section{Construction of Nonlinear Systems from Linear Systems}

\indent\indent
It was shown, that in the classical limit the one-dimensional 
q--oscillator is represented by a reparametrized oscillator~\cite{sol1}. 
The reparametrization is provided by a 
constant of motion and the associated differential equations exhibit 
a special kind of nonlinearity. Here, we would like to consider these 
systems from a general viewpoint: nonlinear systems as ``reparametrized" 
linear ones.

A linear dynamical system, say on $~\hbox {\bf R}^n$, with coordinates
$~\hbox {\bf x}=(x_1,~x_2,~\ldots ,~x_n)~$ is described for any 
n$\times $n matrix $~A~$ by the differential equation
\begin{equation}\label{1}
\dot x_j=A_j^kx_k\,;~~~~~A_j^k\in \hbox {\bf R}\,,
\end{equation}
with solutions
$$
\hbox {\bf x}\,(t) = e^{tA}\hbox {\bf x}_{0}\,;~~~~~\hbox {\bf x}_{0}
 =\hbox {\bf x}\,(t_{0})\,.
$$
To obtain new (nonlinear) systems from the above, we replace
the globally constant matrix $~A~$ with a matrix valued function
 $~B\,(\hbox {\bf x})~$ and write the equation
\begin{equation}\label{3}
\dot x_j = B^k_j(\hbox{\bf x})\,x_k 
\end{equation}
with
$$
\frac{d}{dt} B^{k}_{j} =  x_{n}B^{n}_{m}(\hbox {\bf x})\frac
{\partial }{\partial x_{m}}B^{k}_{j}=0\,.
$$
It means that the matrix elements of the matrix $~B~$ are integrals of 
motion for Eq.~(\ref{3}). For this new system, we can write the 
solutions as
$$
\hbox {\bf x}\,(t)
=\exp \,[tB\,(\hbox {\bf x}_{0})]\,\hbox {\bf x}_{0}\,.
$$
For each initial condition, Eq.~(\ref{3}) reduces to Eq.~(\ref{1}) and
a particular case is the orbit-dependent time reparametrization.

An example of such a deformed equation in field theory has been shown
in~\cite{qdef} where a physical parameter was made dependent on
Cauchy data. A particular family of such systems is when
$$
B^{i}_{j}(\hbox {\bf x}) = f\,(\hbox {\bf x})\,A^{i}_{j}\,,
$$
with $~f~$ any constant of motion of the system.
For instance, let us consider the three-dimensional isotropic
harmonic oscillators
\begin{eqnarray}\label{5}
\dot {\hbox {\bf x}}&=&\omega \hbox {\bf y}\,;\nonumber\\
\\
\dot {\hbox {\bf y}}&=&-\omega \hbox {\bf x}\,;
~~~~\omega \in \hbox {\bf R}\,.\nonumber
\end{eqnarray}
All constants of motion of this system are functions of
$~b_{ij}(x^{i}y^{j}-x^{j}y^{i})~$ and 
$~a_{ij}(y^{i}y^{j}+x^{i}x^{j})\,.$
If we set new frequency $~\Omega =\Omega \,(\,(b_{ij}(x^{i}y^{j}
-x^{j}y^{i}),~a_{ij}(x_{i}x_{j}+y_{i}y_{j})\,)~$ and make the
replacement $~\omega \rightarrow \Omega ~$ in Eq.~(\ref{5}),
we get a nonlinear system. Solution to this system is given by
\begin{equation}\label{6}
\left (\begin{array}{c}
x_{j}(t)\\
y_{j}(t)\end{array}\right )=\left (\begin{array}{clcr}
\cos t\Omega (x_{0},y_{0})&\sin t\Omega (x_{0},y_{0})\\
-\sin t\Omega (x_{0},y_{0})&\cos t\Omega (x_{0},y_{0})\end{array}\right )
\left (\begin{array}{c}
x_{j}(0)\\
y_{j}(0)\end{array}\right )\,.
\end{equation}
For each choice of the functional dependence of $\Omega $ on the constants
of motion, we get a foliation of the carrier space
$\hbox {\bf R}^{6}$ with leaves given by
\begin{equation}\label{7}
\Sigma _{\lambda }=\{(\hbox {\bf x,y})\in \hbox {\bf R}^{6}\,:\,
\Omega \,(\hbox {\bf x,y})=\lambda \}\,.
\end{equation}
For each initial condition 
$~\hbox {\bf x}_{0},\hbox {\bf y}_{0}\in \Sigma _{\lambda }\,,$ 
the oscillatory motion has the same frequency for each one of the 
coordinates.

It is however possible to start with
\begin{equation}\label{8}
\left (\begin{array}{c}
\dot x_{1}\\
\dot y_{1}\\
\dot x_{2}\\
\dot y_{2}\\
\dot x_{3}\\
\dot y_{3}\end{array}\right )
=\left (\matrix{
0&\omega _{1}&0&0&0&0\cr
-\omega _{1}&0&0&0&0&0\cr
0&0&0&\omega _{2}&0&0\cr
0&0&-\omega _{2}&0&0&0\cr
0&0&0&0&0&\omega _{3}\cr
0&0&0&0&-\omega _{3}&0\cr}\right )
\left (\begin{array}{c}
x_{1}\\
y_{1}\\
x_{2}\\
y_{2}\\
x_{3}\\
y_{3}\end{array}\right )\,,
\end{equation}
and ``nonlinearize" the system by making different choices of the 
constants of motion for the different frequencies. For instance, 
we could take
\begin{eqnarray}\label{9}
\Omega _{1}&=&f_{1}(x_{1}^2 + y_{1}^2, x_{2}^2 + y_{2}^2,x_{3}^2 + 
y_{3}^2)\,;\nonumber\\
\Omega _{2}&=&f_{2}(x_{1}^2 + y_{1}^2, x_{2}^2 + y_{2}^2,x_{3}^2 + 
y_{3}^2)\,;\\
\Omega _{3}&=&f_{3}(x_{1}^2 + y_{1}^2, x_{2}^2 + y_{2}^2,x_{3}^2 + 
y_{3}^2)\,.\nonumber
\end{eqnarray}
Now each mode will be ``dynamically coupled" to the others in a
different way.

If $~\Gamma ~$ is the dynamical vectorfield, it should be noticed that 
when we replaced $~\Gamma $ with $~f\Gamma \,,$ i.e., we reparametrize 
$~\Gamma $ by a constant of motion, we get the
{\it same phase portrait for both vectorfields}. If we reparametrize
different modes of vibration differently, we change the phase portrait 
of the new system with respect to the original one of $~\Gamma $. Thus, 
our ``nonlinearization" procedure is more than a parametrization of the 
original linear system.

\section{Deformation of Linear Hamiltonian Systems and Symmetries}

\indent\indent
The above variety of choices has to be restricted if we start with a 
system admitting a Hamiltonian description and want to preserve the
Hamiltonian character after we ``reparametrize" it. The simplest choice 
to get a reparametrized Hamiltonian system is the following:
Start with a linear Hamiltonian system
\begin{equation}\label{10}
\left (\begin{array}{c}
\dot x_{i}\\
\dot y_{i}\end{array}\right )=\left (\begin{array}{c}
\partial H/\partial y_{i}\\
-\partial H/\partial x_{i}\end{array}\right )\,,
\end{equation}
where
$$
H=x_{i}A_{j}^{i}y^{j}+B_{ij}y^{i}y^{j}+C^{ij}x_{i}x_{j}
$$
(the coordinates $~y_i~$ are the momentum components)
and consider the new system associated with the Hamiltonian
$~\widetilde H=f\,(H)\,.$ We get the new nonlinear equations of
motion given by
\begin{eqnarray}\label{11}
\dot x_{i}&=&f'(H)\frac {\partial H}{\partial y_{i}}\,;\nonumber\\
\\
\dot y_{i}&=&-f'(H)\frac {\partial H}{\partial x_{i}}\nonumber
\end{eqnarray}
with $~f'(H)=\partial f/\partial H\,.$
This system can be explicitly integrated as any linear system can be.
This has already been demonstrated in~\cite{sol1}.

To make contact with~\cite{sol1}, we also show, however, a different 
path to nonlinear Hamiltonian systems obtained from linear systems, 
from  linear symmetries to nonlinear ones. If $~A$ is any 
n$\times $n matrix, we denote by $~X_{A}~$ the associated
vectorfield setting
$~X_{A}=x_{i}A_{j}^{i}\,\partial /\partial x_{j}\,.$ For a dynamical
system, we prefer using the notation $~\Gamma ~$ instead of 
$~X_{A}\,.$
The Lie algebra of symmetries for $~\Gamma ~$ contains all linear
vectorfields $~X_{B}~$ such that $~[B,\,A]=0\,;$ this follows trivially
from $~[X_{A},\,X_{B}]=X_{[A,B]}\,.$ In the case of the isotropic harmonic
oscillator, we find that for the $~m$--dimensional oscillator the symmetry
algebra is $~gl\,(m,C)\,.$ Now we ``reparametrize" any one of $~X'_{B}~$
by using constants of motion for $~\Gamma \,.$ Had we started with
matrix $~{\cal B}~$ generating symmetry transformations instead of
infinitesimal ones, i.e., $~{\cal B}^{-1}A{\cal B}=A\,$ (diffeomorphism
in the differential geometric language), we could generate nonlinear
changes of coordinates by following the same idea of ``reparametrization.''
Generally, this procedure will turn canonical transformations into 
nonlinear noncanonical transformations.

We deal with the two-dimensional isotropic harmonic oscillator on
$~\hbox {\bf R}^{4}$
\begin{equation}\label{12}
\dot x_{i} = y_{i}\,;~~~~~
\dot y_{i} = -x_{i}\,;~~~~~i=1,2
\end{equation}
and define the change of coordinates
\begin{eqnarray}\label{13}
q_{i}&=&f_{i}(H_{i})\,x_{i}\,;\nonumber\\
\\
p_{i}&=&f_{i}(H_{i})\,y_{i}\nonumber
\end{eqnarray}
with $~H_{i}=x_{i}^{2}+y_{i}^{2}\,;~~f_{i}\,:\,\hbox {\bf R}\rightarrow
\hbox {\bf R}~$ and no summation on $~i\,.$ In these new coordinates,
the equations of motion have still a linear form given by
\begin{eqnarray}\label{14}
\dot q_{i}&=&p_{i}\,;\nonumber\\
\\
\dot p_{i}&=&-q_{i}\,.\nonumber
\end{eqnarray}
Without loss of generality, we consider only one degree of freedom.
In both coordinate systems, the dynamics admits a Hamiltonian description
given respectively by
\begin{equation}\label{15}
a)~~~~~\{x,y\}=1\,;~~~~~H=\frac {1}{2}\,(x^{2}+y^{2})\,;~~~~~
\frac {d}{dt}f=\{H,f\}
\end{equation}
and
\begin{equation}\label{16}
b)~~~~~\{p,q\}=1\,;~~~~~\widetilde H=\frac {1}{2}\,(p^{2}+q^{2})\,;
~~~~~\frac {d}{dt}f=\{\widetilde H,f\}\,.
\end{equation}
However, using the Poisson bracket of $~a)~$ to compute the Poisson 
bracket in $~b)~$ for $~(p,\,q)~$ as functions of $~(x,\,y)~$ given by 
the system of equations~(\ref{14}), we find
$$
\{f(H)x,\,f(H)y\}= \frac{d}{dH}(Hf^{2}(H)\,)\neq 1\,,
$$
i.e., the nonlinear transformation we have performed is noncanonical. 
The noncanonical property is there even if $~F(H)~$ is a constant 
$~f\,.$ To obtain the same right hand side of the Poisson bracket 
in $~b)\,,$ it is necessary to use for the independent variables 
$~(x,\,y)~$ the new Poisson bracket
\begin{equation}\label{17}
\{x,y\}' = \left [\frac{d}{dH}(Hf^{2}(H)\,)\right ]^{-1}\,;
\end{equation}
with it and the Hamiltonian function
\begin{equation}\label{18}
H'(x,y) = \frac{1}{2}\,(x^{2} + y^{2})~f^{2}(x^{2} + y^{2})\,,
\end{equation}
we have another Hamiltonian description in the coordinate $~(x,\,y)~$ 
for $~\Gamma \,.$

This can be written in symplectic terms as
\begin{equation}\label{19}
i_{\Gamma} \omega = -dH\,;~~~~\omega = dx\wedge dy\,
\end{equation}
and
\begin{equation}\label{20}
i_{\Gamma} \omega' = -dH'\,;~~~~\omega'
= \frac{d}{dH}(Hf^{2}(H)\,)dx\wedge dy\,.
\end{equation}
Because both $~H~$ and $~\widetilde H~$ are Hamiltonian functions for
$~\Gamma \,,$ we can use $~H~$ expressed as function of $~(p,\,q)\,$ 
through the inverse of (\ref{13}) and the bracket $~\{p,\,q\}=1~$ to
introduce a new vectorfield. Otherwise, we can use
$$
\{x, y\}'=\left [\frac{d}{dH}(Hf^{2}(H)\,)\right ]^{-1}
$$
and the Hamiltonian
$$
H=\frac {1}{2}\,(x^{2}+y^{2})\,.
$$ 
In both cases,
we get a ``reparametrized" version of harmonic oscillator. In comparing 
the two procedures, it is clear that in one approach we have used the
same Poisson Brackets (``commutation relations") as before with
a new Hamiltonian which is a function of the standard quadratic 
Hamiltonian. In the second approach, we use the same Hamiltonian 
function but the ``deformed" Poisson brackets (``deformed commutation 
relations"). It is this second viewpoint which allows the connection
with q--oscillators, while the first one has suggested the interpretation
of the classical counterpart as a nonlinear f--oscillator. However,
it should be clear that both views are legitimate and both of them
give rise to a nonlinear dynamics out of a linear one by taking recourse
to a ``reparametrization.''

Having in mind the passage to quantum mechanics, we prefer turning now
to complex variables in $\hbox {\bf R}^{4}$ and defining complex 
coordinates
\begin{eqnarray}\label{CC1}
\alpha_{k} &=&2^{-1/2}(x_{k} + iy_{k})\,;\nonumber\\
\\
\alpha^{*}_{k} &=&2^{-1/2}(x_{k} - iy_{k})\,;~~~~~k = 1,2\,.\nonumber
\end{eqnarray}
The complex structure in $~\hbox {\bf R}^{4}~$ is given by the matrix
$$
J = \left ( \begin{array}{clcr}
0 & 1 & 0 & 0\\
-1 & 0 & 0 & 0\\
0 & 0 & 0 & 1\\
0 & 0 & -1 & 0\end{array}\right )
$$
satisfying $~J^{2}=-1\,.$ This matrix defines a complex structure commuting
with the dynamical evolution associated with the isotropic harmonic 
oscillator
\begin{equation}\label{CC2}
\Gamma = i\,\left (\alpha_{k} \frac {\partial}{\partial \alpha_{k}}
- \alpha^{*}_{k}\frac {\partial}{\partial \alpha^{*}_{k}}\right )\,.
\end{equation}
The equations of motion are
\begin{eqnarray}\label{CC3}
\dot \alpha &=& -i\alpha \,; \nonumber\\
\\
\dot \alpha ^{*}&=&i\alpha ^{*}\,.\nonumber
\end{eqnarray}
For the $~(p,\,q)~$ coordinates, we have new complex coordinates
\begin{eqnarray}\label{CC4}
\xi_{k}&=&2^{-1/2}(q_{k} + i p_{k})\,;\nonumber\\
\\
\xi^{*}_{k} &=&2^{-1/2}(q_{k} - i p_{k})\nonumber
\end{eqnarray}
and, setting
$$
n_{k} = \alpha_{k} \alpha^{*}_{k}\,,
$$
we find
\begin{eqnarray}\label{CC5}
\xi_{k} &=& f_{k}(n_{k})\,\alpha_{k}\,;\nonumber\\
\\
\xi^{*}_{k} &=&f_{k}(n_{k})\,\alpha^{*}_{k}\,.\nonumber
\end{eqnarray}
This transformation is not ``analytic" (we notice that analyticity 
depends on the complex structure and here we have two alternative complex
structures compatible with the dynamics).

The equations of motion for these new variables are the same as 
(\ref{CC3}) as it was seen earlier, i.e., Eqs.~(\ref{CC3}) are form 
invariant under the transformation (\ref{CC5}). In these complex
coordinates, if we take the new point of view, stemming from quantum
mechanics, which takes the Hamiltonian as a primitive concept for the
dynamics, we are naturally led to consider the following two
Hamiltonians                
\begin{equation}\label{CC6}
H_{1} = \frac {1}{2}\,\sum_{k} \alpha_{k}\,\alpha_{k}^{*}
\end{equation}
in the $~\alpha $--coordinates, and
\begin{equation}\label{CC7}
H_{2} = \frac {1}{2}\,\sum_{k} \xi_{k}\,\xi^{*}_{k}
\end{equation}
in the $~\xi$--coordinates. To compare, we express them in the same 
variables to find
\begin{equation}\label{CC8}
H_{1} = \frac {1}{2}\,\sum_{k} n_{k}
\end{equation}
and
\begin{equation}\label{CC9}
H_{2} = \frac {1}{2}\,\sum_{k} f_{k}^{2}(n_{k})\,n_{k}\,.
\end{equation}
We now use the same bracket for them, say
\begin{equation}\label{CC10}
\{\alpha_{k}, \alpha^{*}_{j}\} = -i \delta_{kj}\,.
\end{equation}
We obtain two different dynamical systems, where the one associated
with $~H_{2}$ is not even necessarily isotropic. The evolution goes
from a periodic orbit to an orbit whose closure is a two-dimensional
torus, i.e., the associated systems are completely different. Of course, 
there is no contradiction. Indeed, to have the same dynamics using 
$~H_{2}~$ we should use different Poisson brackets, as it has been 
seen earlier.

For the nonlinear oscillator obtained by means of the deformation
function $~f,$ the equations of motion are
\begin{eqnarray}\label{CC11}
\dot \alpha &=& - i\,\frac {d}{dn}(nf^{2}(n)\,)\,\alpha \,;
\nonumber\\
\\
\dot \alpha^{*} &=&i\,\frac{d}{dn}(nf^{2}(n)\,)\,\alpha^{*}\nonumber
\end{eqnarray}
and are not invariant under the transformation (\ref{CC5}), which indeed
gives the equations of motion for another system of our class but
with a different deformation function. They admit different 
Hamiltonian descriptions. For instance,
\begin{eqnarray}\label{CC12}
 &&1.~~H = \alpha \alpha^{*}\,;~~~~~~~~\{\alpha, \alpha^{*}\} = i\,\frac
{d}{dn}(nf^{2}(n)\,)\,;~~~~\omega = \left (\frac{d}{dn} (nf^{2}(n)\,)
\right )^{-1} d\alpha \wedge d\alpha^{*},\nonumber\\[6pt]
\\
&& 2.~~H =  nf^{2}(n)\,;~~~\{\alpha, \alpha^{*}\} = - i\,;~~~~~~~~~~
\omega = d\alpha\wedge \alpha^{*}.\nonumber
\end{eqnarray}
It is now clear, that in the quantum picture these complex coordinates
will be replaced by creation and annihilation operators. Therefore for 
the corresponding commutators, we can repeat what we have said for the
Poisson bracket.

\section{Some Examples}

\indent\indent
The first example we consider concerns the classical one-dimensional 
q--oscillator \cite{sol1}, where
\begin{eqnarray}\label{CC13}
\xi &=& \sqrt{\frac{\sinh \,\lambda \alpha \alpha^*}{\alpha \alpha^* \,
 \sinh \,\lambda}} \alpha \,;\nonumber\\
\\
\xi^* &=& \sqrt{\frac{\sinh \, \lambda \alpha \alpha^*}{\alpha
 \alpha^* \, \sinh \,\lambda}} \alpha^*\,.\nonumber
\end{eqnarray}
The Poisson bracket for the new variables can be computed using 
(\ref{CC10}) and expressed in terms of themselves, by the use of the map 
$~(\xi ,\xi ^{*})\rightarrow (\alpha ,\alpha ^{*})\,,$ the inverse of 
(\ref{CC5}) (which in this case exists). One then obtains
\begin{equation}\label{CC14}
\{ \xi, \, \xi^*\} = -i\,\frac { \lambda}{\sinh \,\lambda}\,\sqrt{1 +
 |\xi|^{4}(\sinh \,\lambda)^2}\,,
\end{equation}
so that we can consider a new system described by such variables with
Hamiltonian function
\begin{equation}\label{CC15}
H(\xi,\xi^*) =  \xi\xi^*.
\end{equation}
The equations of motion then are
\begin{equation}\label{CC16}
\dot \xi = - i\,\frac {\lambda}{\sinh \,\lambda} \,\sqrt {1 + |\xi|^4
(\sinh \,\lambda )^2\,}\xi 
\end{equation}
(and its complex conjugate) with solutions
\begin{equation}\label{CC17}
\xi \,(t) = \xi \,(0) \exp \,\left [\frac {-it  \lambda}{\sinh \,\lambda}\,
\sqrt {1 + \mid \xi \,(0)\mid ^{4}(\sinh \,\lambda)^2\,}\right ]
\end{equation}
(and its conjugate). Such a system can be rewritten, of course,
in terms of $~(\alpha,\,\alpha^*)~$ variables: in these coordinates,
the original Poisson bracket is unchanged while the Hamiltonian
function is 
\begin{equation}\label{CC18}
H'(\alpha, \alpha^*)=\frac {\sinh \,\lambda \alpha \alpha^*}
{\sinh \,\lambda}\,.
\end{equation}
It is clear, that this new dynamical system has a phase portrait which
is the same as the usual linear  harmonic oscillator. The only difference 
is in the frequencies, the new one being
\begin{equation}\label{CC19}
\omega  
=\frac {\lambda }{\sinh \,\lambda} \, \cosh \,\lambda \alpha \alpha ^* \,.
\end{equation}
We notice that $~\alpha \alpha^*~$ is a constant of motion for both 
systems. The deformation function used has given both energy and frequency
exponentially growing with $~\alpha \alpha^*.$ To have physically
more acceptable functions, it is not difficult to consider a slightly
different deformation function.

The other example we consider is the deformation which leads to the 
classical version of harmonious states to be discussed in next 
sections. Here the deformation function is taken to be
\begin{equation}\label{CC20}
f(n) = \left(\frac{1}{\Gamma(n+1)}\right)^{1/2}\,;~~~ n 
= \alpha \alpha^{*},
\end{equation}
so that
\begin{equation}\label{CC21}
\xi = \left(\frac{1}{\Gamma(n+1)}\right)^{1/2} \alpha \,;~~~~~~
\xi^{*} = \left(\frac{1}{\Gamma(n+1)}\right)^{1/2} \alpha^{*}.
\end{equation}
The Poisson bracket
\begin{equation}\label{CC22}
\{\xi, \xi^{*}\} = \frac {d}{dn}\left(\frac {n}{\Gamma(n+1)}\right),
\end{equation}
where the right hand side is the function of $~\xi \xi^{*}~$ which is 
its transform by the inverse of (\ref{CC5}), remarking that the 
$~\Gamma$ function on the positive real line has well defined derivative 
as well as inverse.

\section{Nonlinear Oscillators in Quantum Mechanics through 
Noncanonical Transformations}

\indent\indent
We wish now to deal with the quantum analogue systems: we will
remain in the usual formulation where the Fock space is better to 
describe the quantum harmonic oscillator. To go along the same 
lines as in the previous sections, we stay in the Heisenberg picture 
and write the equations of motion for the harmonic oscillator amplitude 
$~a$
$$
\dot a =- i \omega a
$$
and for its conjugate
$$
\dot a^{\dagger} = i \omega a^{\dagger}\,.
$$
The transformation~(\ref{CC5}) is written here as
$~A =a\,f(a^{\dagger}a)\,;~A^{\dagger} 
= f(a^{\dagger}a)\,a^{\dagger}\,.$
It is noncanonical since it does not preserve the commutation
relations. The operators $~A$ and $~A^{\dagger}$ evolve with the same 
equations, i.e.,
$$
\dot A=-i\omega A\,,~~~~~\dot A^{\dagger} =i\omega A^{\dagger} 
$$
in complete analogy with the classsical case. We have in our Hilbert 
space the vacuum state $~\mid 0\rangle ~$ which satisfies
$$
a\mid 0\rangle =0\,,
$$ 
as well as 
$$
A\mid 0\rangle =0\,.
$$
This allows us to construct two bases in the vector space having this
vector in common. One is the standard (Fock) basis
$$
\mid n\rangle =\frac {a^{\dagger n}}{\sqrt {n!}}\mid 0\rangle \,,
$$
which is orthonormal in standard scalar product
$$
\langle n\mid m\rangle =\delta _{nm}\,.
$$
Another basis is constructed using the operator $~A^{\dagger},$
$$
\mid \widetilde n\rangle =\frac {A^{\dagger n}}{\sqrt {n!}}\mid 0
\rangle .
$$
We define a {\it new} scalar product in the same vector space which  
gives 
$$
\langle \widetilde n\mid \widetilde m\rangle =\delta _{nm}\,.
$$
Provided $~f(a^{\dagger}\,a)~$ to be nonsigular, we can then speak of 
two Hilbert space structures carried by the same vector space.

The adjoint with respect to this new scalar product does not coincide
with the old one. We can then define the operators
$$
b^{*}\mid \widetilde n\rangle =\sqrt {n+1}\mid \widetilde {n+1}
\rangle \,;~~~~~b\mid \widetilde n\rangle =\sqrt {n}\mid 
\widetilde {n-1}\rangle \,;
$$
where $~*~$ means the adjoint in the new scalar product. These 
operators satisfy the commutation relations $~[b,\,b^{*}]=1\,.$ 
Taking the Hamiltonian $~H=\omega b^{*}b$ we have for the operators 
$~b,\,b^{*}~$ the equation of motion of the harmonic oscillator. 
This is actually the situation with $~A~$ and $~A^{\dagger}.$

In the second Hilbert space, the operators $~A~$ and $~A^{\dagger}~$ 
have an identical representation as $~a~$ and $~a^{\dagger}~$ have
in the Fock space and so they satisfy the commutation relation
$~[A,\,A^{\dagger}] = 1\,.$
Thus, for one and the same vector space, we have the possibility to 
introduce two Hilbert space structures. As for the dynamics of the 
harmonic oscillator, we have two different descriptions, 
which parallel the alternative Hamiltonian descriptions of the 
classical oscillator. Therefore, much as we did for the classical case, 
we can use the new Hamiltonian and the old commutation relations to get 
a ``deformed" dynamics, or vice versa. To keep the current physical 
interpretation of the operators $~a$ and $~a^{\dagger}$, we 
choose to maintain the value of their commutator which is directly 
connected with  measurements, while we will consider deformed Hamiltonian 
operators. Then operator of ``energy" $~\widetilde {H}~$ is
\begin{equation}\label{NO1}
\widetilde {H} = \frac {\omega}{2}\,(A^{\dagger}A + AA^{\dagger})\,.
\end{equation}
In the Fock space, its eigenvalues are
\begin{equation}\label{NO2}
E_{n}=\frac {\omega}{2}\,\left [\,(n+1)\,f(n+1)\,f^{*}(n+1) 
+ n\,f(n)\,f^{*}(n)\,\right ]\,.
\end{equation}
We illustrate the situation with an example and consider the
nonlinear noncanonical transformation such that
\begin{equation}\label{NO3}
 A = a f_{q}(a^{\dagger}a) = a\,\sqrt {\frac {\sinh \,\lambda\hat {n}}
{\hat{n}\sinh \,\lambda}}\,,
\end{equation}
which is invertible,
\begin{equation}\label{1x}
a=A\left[\frac 
{\ln \,\left [\hat{N}\sinh \,\lambda 
+\sqrt {\hat{N}^{2}\sinh \,^{2}\lambda +1\,}\,\right ]}
{\lambda \hat{N}} \right]^{1/2},
\end{equation}
where
\begin{equation}\label{2x}
\hat {N}=A^{\dag }A\,.
\end{equation}
The operators $~A,\,A^{\dag }~$ acting in the same Hilbert space as the
operators $~a,\,a^{\dag }~$ (the original Fock space) satisfy in this 
case the commutation relations
\begin{equation}\label{3x}
[A,A^{\dag }]=\hat{N}(\cosh \lambda -1)+\sqrt {\hat{N}^{2}
\sinh \,^{2}\lambda +1}\,,
\end{equation}
as can be seen expressing all the operators in terms of matrices.

If one has the linear dynamics for the operator $~a~$ with frequency
equal to unity, i.e.,
\begin{equation}\label{4x}
\dot a+ia=0\,,
\end{equation}
and boson commutation relation for the operators $~a~$ and 
$~a^\dagger ,$ the same dynamics exists for the operator $~A\,:$
\begin{equation}\label{6x}
\dot A+iA=0\,,
\end{equation}
since
\begin{equation}\label{7x}
\dot A+iA = (\dot a + ia) f_{q}(a^{\dagger}a)
\end{equation}
and
the function $~f_{q}~$ is an integral of motion. The Hamiltonian for 
this dynamics may be taken as
\begin{equation}\label{8x}
H=\frac {1}{\lambda }\ln \left [A^{\dag }A\sinh \,\lambda
+\sqrt {(A^{\dag }A)^{2}\sinh \,^{2}\lambda +1}\,\right ]+\frac {1}{2}\,,
\end{equation}
and one obtains
\begin{equation}\label{9x}
[A, H] = A\,.
\end{equation}
For the same dynamics, it is possible however to have a different 
Hamiltonian formulation. In one case, it is related to the above 
Hamiltonian, while in another Hilbert space we define the Hamiltonian
\begin{equation}\label{10x}
H'=A^{\dag }A+\frac {1}{2}
\end{equation}
with commuation relation
\begin{equation}\label{11x}
[A,A^{\dag }]=1\,,
\end{equation}
and again
\begin{equation}\label{12x}
[A, H'] = A\,.
\end{equation}
Thus, we see that analogous to classical mechanics there are possible 
alternative descriptions of a quantum system. For the same equations 
of motion, for the operators we have two different Hamiltonians 
with corresponding different commutation relations.

The same situation is anyhow present also for the class of nonlinear 
oscillators we have considered. There, continuing with the example 
of the $f_{q}$ deforming function, we start with the dynamics 
for the q--oscillator \cite{sol1} given by the equation
\begin{equation}\label{13x}
\dot a+i\,\frac{1} {2\,\sinh \,\lambda }\,[\,\sinh \,\lambda (a^{\dagger}a 
+2)-\sinh \,\lambda a^{\dagger}a\,]\,a = 0\,.
\end{equation}
It worth remarking at this point, that if we multiply the last equation 
from the right hand side by the same function $~f_{q}(a^{\dagger}a)\,,$
for instance, we are led to a new f--oscillator. Since this
function is an integral of motion for the above q--nonlinear equation,
we have in fact for operator~(\ref{NO3})
the equation of motion
\begin{equation}\label{15x}
\dot A+i\,\frac {\sinh \,\lambda }{\lambda }(\cosh \,\lambda a^{\dag }a)A
=0\,.
\end{equation}
But after (\ref{1x}),
\begin{equation}\label{16x}
a^{\dag }a=\frac {1}{\lambda }\,\ln \left [A^{\dag }A\sinh \,\lambda
+\sqrt {(A^{\dag }A)^{2}\sinh \,^{2}\lambda +1}\,\right ],
\end{equation}
the obtained dynamics is different from the initial one.

We return now to consider Eq.~(\ref{13x}) and the first Hamiltonian
description is given by
\begin{equation}\label{17x}
H = \frac{1} {2\sinh \,\lambda }\,[\,\sinh \,\lambda (a^{\dagger}a + 1) +
\sinh \,\lambda a^{\dagger}a\,]\,;~~~~~~[a, a^{\dagger}] = 1\,.
\end{equation}
In another Hilbert space, let the operators $~B$ and $~B^{\dagger}$
evolve with the equation
\begin{equation}\label{18x}
\dot B+i\,\frac{1} {2\sinh \,\lambda }\,[\,\sinh \,\lambda (B^{\dagger}B 
+2)-\sinh \,\lambda B^{\dagger}B\,]\, B = 0\,.
\end{equation}
If we take the commutation relation
\begin{equation}\label{19x}
[B,B^{\dag }]=B^{\dag }B(\cosh \lambda -1)
+\sqrt {(B^{\dag }B)^{2}\sinh \,^{2}\lambda +1}\,,
\end{equation}
the form of the Hamiltonian for this system differs from
the form of Hamiltonian~(\ref{17x}).

The important physical consequence of the existence of the same dynamics
for quadratures with different commutation relations is the possibility
of existing identical harmonic vibrations of two kinds. One vibrational 
process respects the Heisenberg uncertainty relation since quadratures
satisfy standard boson commutation relations. Another vibrational process
is compatible with different uncertainty relation for its quadratures since
they satisfy different commutation relations. Nevertheless, from the view
point of dynamics (harmonic vibrations) both cases are undistinguishable. 
                                     
\section{f-oscillator Operators}

\indent\indent
The operators $~A~$ and $~A^{\dagger}~$ represent the dynamical variables
to be associated with the quantum f--oscillators. The well known 
q--oscillator operators belong to this class. In this section,
after discussing some algebraic features useful for the 
f--generalization, we refer also to other examples of f--oscillators 
already taken into consideration.

We start by recalling some notions about the harmonic oscillator
operators $~a~$ and $~a^{\dagger}~$ whose algebraic structure is
$~[a,\,a^{\dagger}]=1\,.$ In the Fock space with
$~a = (a^{\dagger})^{\dagger}\,;~\hat n = a^{\dagger}a\,,$
the basis is given by the eigenfunctions of $~\hat n$
\begin{equation}\label{AR3}
\hat n\mid n\rangle =n\mid n\rangle \,;~~~~~n \in \hbox {\bf Z}^{+}\,.
\end{equation}
We have also
\begin{equation}\label{AR5}
{\hbox {\bf 1}} = \sum_{0}^{\infty} \mid n\rangle \langle n\mid \,;
~~~~~\langle n\mid m\rangle = \delta_{nm}
\end{equation}
and $~\forall~ {f\,:\,\hbox {\bf Z}^{+}} \rightarrow \hbox {\bf C}$
\begin{equation}\label{AR6}
f\,(\hat n) = \sum_{j=0}^{\infty} f(j)\mid j\rangle \langle j\mid \,.
\end{equation}
Consider now a ``distortion" of $~a~$ and $~a^{\dagger}~$ of the form
\begin{eqnarray}\label{AR7}
A &=& a\,f(\hat n) = f(\hat n + 1)\,a\,;\nonumber\\
\\
A^{\dagger} &=& f^{\dagger}(\hat n)\,a^{\dagger} 
= a^{\dagger} f^{\dagger}(\hat n  + 1)\nonumber
\end{eqnarray}
and note that
\begin{equation}\label{AR8}
[A, \hat {n}] = A\,;~~~~~[A^{\dagger}, \hat {n}] = - A^{\dagger}.
\end{equation}
The functions we are considering can be made dependent in general, 
also on continuous parameters, in such a way that for particular 
values of them the usual annihilation and creation operators are 
reconstructed. We will say then that we are in presence of continuous 
deformations. This was the case of q--deformations \cite{polych}. In 
principle, one may consider discontinuous deformations, too.

Since
\begin{eqnarray}\label{AR9}
a&=&\sum_{n=0}^{\infty} \sqrt {n}\mid n-1\rangle \langle n\mid \,;
\nonumber\\
\\
a^{\dagger}&=&\sum_{n=0}^{\infty} \sqrt {n}\mid n\rangle \langle n
-1\mid \,,\nonumber
\end{eqnarray}
the same Fock space is a carrier space for $~A~$ and $~A^{\dagger},$
i.e.,
\begin{eqnarray}\label{AR10}
A&=&\sum_{n=0}^{\infty}\,\sqrt {n} \,f(n)\mid n-1\rangle \langle n
\mid \,;\nonumber\\
\\
A^{\dagger}&=&\sum_{0}^{\infty} \,\sqrt {n}\,f^{*}(n) \mid n\rangle
\langle n-1\mid \,.\nonumber
\end{eqnarray}
This realization may or may not be irreducible depending on the assumed
functional form of $~f\,(n)\,.$

Following the choice of not deforming the commutators of the physical 
variables $~(a,\,a^{\dagger})\,,$ the commutator between $~A~$ and 
$~A^{\dagger}~$ can be easily computed and by using (\ref{AR10}) reads
\begin{equation}\label{AR11}
F(\hat {n}) \doteq{[A,A^{\dagger}]}
= (\hat {n}+1)\,f(\hat {n}+1)\,(\,f(\hat {n}+1)\,)^{\dagger}
 - \hat {n}\,f(\hat {n})\,(\,f(\hat {n})\,)^{\dagger}\,,
\end{equation}
while the q--commutator is
\begin{eqnarray}\label{AR12}
G(\hat n) &\doteq &[A, A^{\dagger}]_{q} \doteq AA^{\dagger} 
- q A^{\dagger}A\nonumber\\
&=&(\hat {n}+1)\,f(\hat {n}+1)\,(f(\hat {n}+1)\,)^{\dagger} 
- q\,\hat {n}\,f(\hat {n})\,(f(\hat {n})\,)^{\dagger}\,;
~~~~q \in (0,1]\,.
\end{eqnarray}
We have, also
\begin{equation}\label{AR13}
F(\hat {n})-G(\hat {n})=(q-1)\,\hat {n}\,f(\hat {n})\,
f^{\dagger}(\hat {n})\,.
\end{equation}
Since only $~ff^{\dagger}$ occurs, the phase of $~f$ is irrelevant and 
we may, without loss of generality, choose $~f$ to be real and 
nonnegative:
\begin{equation}\label{AR14}
f^{\dagger}(\hat {n}) = f\,(\hat {n})\,.
\end{equation}
Alternately, given the functions $~F$ or $~G$, assumed hermitian, we 
obtain the following solution to Eqs. (\ref{AR11}) and (\ref{AR12})
\begin{equation}\label{AR15}
f\,(n) =  \frac {1}{\sqrt n}\,\left (\sum_{j=0}^{n-1} F\,(j)\right )^{1/2}\,;
~~~ n \neq  0
\end{equation}
and
\begin{equation}\label{AR16}
f\,(n)
=\frac {1}{\sqrt n}\left(\sum_{j=0}^{n-1}q^{j}\,G\,(n-j-1)\right)^{1/2}\,;
~~~~n\neq  0\,,
\end{equation}
respectively, with $~f\,(0)~$ arbitrary in both cases. Such solutions
are unique, having been obtained by construction. Of course, 
we may use complex $~f$ to construct the functions $~F~$ and $~G~$
in (\ref{AR11}) and (\ref{AR12}). However, to obtain $~f~$ from $~F~$ 
and $~G$, we remark that $~F~$ has to be real as well as $~G~$ when 
$~q~$ is real. Then in (\ref{AR7}), we have 
the freedom of choosing an $~n~$ dependent phase of $~f\,,$ which
corresponds to construction of generalized coherent 
states~\cite{glat,zof}. These generalized coherent states were 
analyzed in~\cite{spirid}, were dyfferent types of interesting new 
states were introduced.

As was mentioned earlier, in the case of the q--oscillator operators,
the function $~f~$ depends also on a continuous parameter in order to 
obtain the harmonic oscillator operators as a limiting case. 
Starting with the q--commutation relation~\cite{bie,mc}
\begin{equation}\label{qq1}
G\,(\hat n) = q^{-\hat n}\,;~~~~\lambda =\ln q\,;
~~~~\lambda\in \hbox {\bf R}\,,
\end{equation}
using (\ref{AR16}) one obtains
\begin{equation}\label{qq2}
f\,(n) = \sqrt {\frac{1}{n}\,\frac {q^{n} - q^{-n}}{q - q^{-1}}} =
\sqrt {\frac {\sinh \,\lambda n}{n \,\sinh \,\lambda}} 
\doteq f_{q}(n)\,,
\end{equation}
setting
\begin{equation}\label{qq3}
f_{q}(0) = 1\,.
\end{equation}
We see from (\ref{17x}) that the eigenvalues of the ``energy" 
operator grow exponentially with increasing occupation number.

One can remark that the phase operators $~V,\,V^{\dagger}$~\cite{suss} 
are actually deformations of the Bose operators of the kind we are 
studying and lead to the harmonious states~\cite{sud3} 
to be considered below. In this case,
\begin{equation}\label{qq15}
f\,(n) = \frac {1}{\sqrt n} \doteq f_{h}\,,
\end{equation}
so that
\begin{equation}\label{qq16}
A~\mid n\rangle =\mid n-1\rangle \,;~~~~ A^{\dagger}\mid n\rangle
=\mid n+1\rangle \,;~~~~n\neq 0\,;~~~~A\mid 0\rangle =0\,.
\end{equation}

When many degrees of freedom are involved, 
we have two possible choices. The first one defines q--oscillators which 
satisfy q--commutation relations among a single degree of freedom, but 
mutually commuting between different degrees of freedom, possibly with 
different functions for the various degrees of freedom. Another choice 
is to make $~f~$ dependent on the total occupation number operator
\begin{equation}\label{qq17}
\hat n_{\mbox {tot}} = \sum_{i}\,\hat {n}_{i}\,.
\end{equation}
For two degrees of freedom, this was discussed in~\cite{sol1}.

\section{Nonlinear Coherent States}
\indent\indent
Coherent states were originally introduced as eigenstates of the 
annihilation operator for the harmonic oscillator and then widely 
used in physics, particularly in quantum optics. This is therefore 
a concept of algebraic origin and having now constructed a similar 
annihilation operator it is natural, following the same procedure, 
to construct a new class of f--coherent states in the Fock space. 
This construction is in general different from other ones
\cite{ksmmb}.  Further the f--coherent states may not be preserved 
under time evolution.  Nevertheless, we are willing to call them
f--coherent states for an easy identification, of the kind already 
proposed for the eigenstates of the q--annihilation operator which 
were named q--coherent states~\cite{bie,breg}.

Let us take for the one-mode case the operator $~A~$~(\ref{AR7})\,.
Then one can consider the eigenfunctions $~\mid \alpha, \,f\rangle ~$
of $~A~$ in a Hilbert space. They therefore satisfy the equation
\begin{equation}\label{FC3}
A\mid \alpha,\,f\rangle = \alpha \mid \alpha,\,f\rangle \,; ~~~~~
\alpha \in \hbox{\bf C}\,.
\end{equation}
Looking for the decomposition of $~\mid \alpha,\,f \rangle~$ in the 
Fock space
\begin{equation}\label{FC4}
\mid \alpha,\,f\rangle = \sum_{n=0}^{\infty} c_{n} \mid n\rangle \,,
\end{equation}
we obtain for the coefficients $~c_{n}~$ the following recurrence 
relation
\begin{equation}\label{FC6}
c_{n+1}\sqrt {n+1}\,f(n+1) = c_{n}  \alpha \,.
\end{equation}
This gives
\begin{equation}\label{FC7}
c_{n} = c_{0} \,\frac {\alpha^{n}}{\sqrt {n!}\, [f(n)]!}\,,
\end{equation}
in which
\begin{equation}\label{FC8}
[f\,(n)]! = f\,(0)f\,(1)\cdots f\,(n)\,.
\end{equation}
To fix $~c_{0}\,,$ we use the condition
\begin{equation}\label{FC9}
\langle \alpha,\,f\mid \alpha,\,f\rangle = 1
\end{equation}
and obtain
\begin{equation}\label{FC10}
c_{0}=\left(\sum_{n=0}^{\infty }
\frac {|\alpha|^{2n}}{n!\,|\,[f(n)]!\,|^{2}}\right)^{-1/2}\,.
\end{equation}
To emphasize the dependence of $~c_{0}~$ on $~f~$ and 
$~|\alpha |\,,$ we will write
\begin{equation}\label{FC11}
c_{0} = N_{f,\,\alpha}
\end{equation}
and in order to have states belonging to the Fock space it is 
required that
\begin{equation}\label{FC12}
0<N_{f,\,\alpha }<\infty \,,
\end{equation}
therefore not any $~f~$ and $~|\alpha |~$ are allowed.
We will denote with $~\bar \rho ~$ the positive number
such that, given $~f\,,$ the above series converge 
$~\forall \,|\alpha |\,\leq \bar \rho \,.$ The scalar product 
is easily written
\begin{equation}\label{FC13}
\langle \alpha \mid \beta \rangle = N_{f,\,\alpha}~N_{f,\,\beta} \,
\sum_{n=0}^{\infty} \,\frac {1}{n!\,|\,[f(n)]!\,|^{2}}\,
(\alpha^{*}\beta)^{n}\,;~~~~|\alpha |, \,|\beta |
<\bar \rho \,.
\end{equation}
No further constraints are then put on $~f~$ and $~\bar \rho \,.$

It should be remarked furthermore that, given $~C\,(n)=~C_{n}~$
any real function on $~\hbox {\bf Z}^{+}\,,$ the state 
$~\mid \alpha ,\,C\rangle ~$ defined by
\begin{equation}\label{FC14}
\mid \alpha ,\,C\rangle = \sum_{n=0}^{\infty}\,C_{n}\,\alpha^{n}
\mid n\rangle
\end{equation}
is an eigenfunction of some $~A\,.$ In fact, the corresponding
function $~f~$ is found to be
\begin{equation}\label{FC15}
f\,(n) = \frac {1}{\sqrt n}\,\frac{C_{n-1}}{C_{n}}\,.
\end{equation}
Such eigenstate can be normalized if the $~f~$ so obtained satisfies 
(\ref{FC10}). In the case
\begin{equation}\label{FC16}
f\,(n) = 1\,,
\end{equation}
$\mid \alpha,\,1\rangle ~$ denotes the usual coherent state and 
\begin{equation}\label{FC17}
N_{1,\,\alpha} = \exp \left (-\frac {|\alpha |^{2}}{2}
\right )\,,
\end{equation}
$\alpha ~$ can be any complex number.

As anticipated, the known q--coherent states \cite{bie,breg} 
turn out to be a particular case of f--coherent states, which we 
might also call f$_{q}$--coherent states. Normalization factor of 
such states is
\begin{equation}\label{FQ18}
N_{q,\,\alpha }
=\left( \sum_{n=0}^{\infty }\frac {|\alpha|^{2n}}{[n]!}\right)^{-1}\,,
\end{equation}
in which
\begin{equation}\label{FQ19}
[n]!= 
\frac {\sinh \,\lambda n}{\sinh \,\lambda}\,\frac {\sinh \,\lambda(n-1)}
{\sinh \,\lambda}\cdots 1\,.
\end{equation}
Using the notation (\ref{FC8}) we can also write
\begin{equation}\label{FQ20}
[n]! = [n\,f^{2}_{q}(n)]!\,.
\end{equation}
It is seen that $~\alpha $ can be any complex number.
For the scalar product, we have
\begin{equation}\label{FQ21}
\langle \alpha \mid \beta \rangle 
=\left(\sum_{n=0}^{\infty}\frac {|\alpha |^{2n}}{[n]!}\right)^{-1}
\left(\sum_{n=0}^{\infty}\frac {|\beta |^{2n}}{[n]!}\right)
\sum _{n=0}^{\infty }\frac{1}{[n]!}(\alpha ^{*}\beta )^{n}\,.
\end{equation}

Harmonious states~\cite{sud3} are eigenstates of the annihilation operator
deformed by the factor $~f_{h}$~(\ref{qq15}), to which corresponds the 
normalization
\begin{equation}\label{FQ22}
N_{h,\,\alpha} = (1 - |\alpha|^{2})^{-1/2}\,,
\end{equation}
and the acceptable $~\alpha$ must have modulus less the 1\,.
Following (\ref{FC3}) they can be denoted 
$~\mid \alpha, \,f_{h}\rangle ~$ and their scalar product is
\begin{equation}\label{FQ23}
\langle \alpha, \,f_{h}\mid \beta, \,f_{h}\rangle 
=\left(\frac {1}{1-|\alpha|^{2}}\right)^{-1/2} \left(\frac {1}
{1-|\beta|^{2}}\right)^{-1/2} (1-\alpha^{*}\beta)^{-1}\,.
\end{equation}

\section{Nonlinear Coherent States in Different Representations}

\indent\indent
Since the state $~\mid \alpha, \,f\rangle ~$ is given as series of 
Fock states, we can easily write the wave function of these states 
in different representations explicitly.

In coordinate representation, the wave function is
\begin{equation}\label{FQ24}
\psi_{\alpha ,\,f}^{(x)} = {\pi}^{-1/4} N_{\alpha, \,f}\, 
e^{-x^{2}/2}\,\sum_{n=0}^{\infty}\left(\frac {\alpha}{\sqrt 2}
\right)^{n}\,\frac {1}{n!\,[f(n)]!}\,H_{n}(x)\,,
\end{equation}
where $~H_{n}~$ is the Hermite polynomial of degree $~n\,.$

For the momentum representation, the formula is the same.

For the Bargmann representation (the usual coherent states), 
the wave function $~\langle z\mid \alpha ,\,f\rangle \,,$ where
we use the basis $~\mid z\rangle \,\,(z\in \hbox{\bf C})~$ with 
$~a\mid z\rangle  = z\mid z\rangle \,,$ takes the form
\begin{equation}\label{FQ25}
\psi_{\alpha ,\,f}^{(z)} = N_{\alpha, \,f}\,e^{-|z|^{2}/2}\,
\sum_{n=0}^{\infty} (z^{*}\alpha)^{n}\,\frac {1}{n!\,[f(n)]!}\,.
\end{equation}
For a continuous parameter $~f\,,$ in the limit
$~f\rightarrow 1$ the usual wave function is recovered
\begin{equation}\label{FQ26}
\psi_{\alpha ,\,1}^{(z)}=\exp \,{\left (\frac {|z|^2-|\alpha|^2}{2} 
+z^{*}\alpha\right )}\,.
\end{equation}

In the Wigner--Moyal representation~\cite{wig}, the density matrix
for the f--coherent state reads
\begin{equation}\label{FQ27}
W_{f}(x,p)= 2\,N_{\alpha, \,f}^{2}\,e^{-(x^2 + p^2)}\sum_{m=0}^{\infty}
\sum_{n=0}^{\infty} \frac {1}{m!\,[f(m)]!}\frac {1}{[f(n)]!}
\alpha ^n(-\alpha ^*)^m [\sqrt {2}(x-ip)]^{m-n} L_{n}^{m-n}(2(x^2 +
p^2))\,,
\end{equation}
where $~L_{m}^{n}~$ denotes a generalized Laguerre polynomial.
For the particular case of q--oscillator $~f = f_{q}\,,$
\begin{equation}\label{FQ28}
W_{q,\,\alpha}=2\,
\left( \sum_{0}^{\infty}\frac {|\alpha|^{2n}}{[n]!}\right)^{-1}
e^{-(x^2 + p^2)}\sum_{m=0}^{\infty}
\sum_{n=0}^{\infty} \frac {1}{[m]!}\frac {n!}{[n]!}\alpha ^n 
(-\alpha ^*)^m [\sqrt {2}(x-ip)]^{m-n} L_{n}^{m-n}(2(x^2 +p^2 ))\,.
\end{equation}

Finally, we consider Husimi--Kano~\cite{hus} Q--function of 
f--coherent states. In the definition, $~Q_{\psi}(z, z^*)~$ 
is the diagonal matrix element of the density operator 
$~\mid \psi \rangle \langle \psi \mid ~$
for the state $~\mid \psi \rangle ~$ 
in the usual coherent state basis. For an f--coherent state 
$~\mid \alpha, \,f\rangle \,,$ we can write
\begin{equation}\label{FQ29}
Q_{f}(z,\,z^*) = e^{-|z|^2} N_{f, \,\alpha}^{2} \sum_{m=0}^{\infty} 
\sum_{n=0}^{\infty} \frac {(z^{*}\alpha)^{m}}{m!\,[f(m)]!}
\,\frac {(z\alpha ^{*})^{n}}{n!\,[f(n)]!}\,.
\end{equation}
For q--coherent state, the Q--function is
\begin{equation}\label{FQ30}
Q_{q}(z,\,z^*)=e^{-|z|^2}\left(\sum_{0}^{\infty}
\frac {|\alpha|^{2n}}{[n]!}\right)^{-2} \sum_{m=0}^{\infty} 
\sum_{n=0}^{\infty}\,\frac {(z^*\alpha )^{m}}{\sqrt{m![m]!}}
\frac {(z\alpha ^*)^{n}}
{\sqrt{n![n]!}}\,.
\end{equation}
For the harmonious states, the Wigner--Moyal function is
\begin{equation}\label{FQ31}
W_{h}(x,\,p) = 2\,(1-|\alpha|^2)e^{-(x^2 + p^2)}\sum_{m=0}^{\infty}
 \sum_{n=0}^{\infty} \sqrt{\frac {n!}{m!}}\,
\alpha ^n 
(-\alpha ^*)^m [\sqrt {2}(x-ip)]^{m-n} 
L_{n}^{m-n}(2(x^2 +p^2))
\end{equation}
and the Husimi--Kano function
\begin{equation}\label{FQ32}
Q_{h}(z,\,z^*) = e^{-|z|^2} (1-|\alpha|^2) \sum_{m=0}^{\infty} 
\sum_{n=0}^{\infty} \frac {(z^{*}\alpha)^{m}}
{\sqrt{m!}}\frac {(z\alpha ^{*})^{n}}{\sqrt{n!}}\,.
\end{equation}

\section{f-Coherent State Evolution}

\indent\indent
Here we offer a few considerations with the aim of underlining the
peculiarities of the new classes of states in the Fock space, 
cautioning against relying too much  on the  abuse of their name.

We consider our field mode evolution to be guided by the equation
of motion with the quantum Hamiltonian
\begin{equation}\label{FCE1}
\widetilde {H}=\frac {1}{2}\,(A^{\dagger} A+A  A^{\dagger})\,,
\end{equation}
i.e., in the variables $~(a,\,a^{\dagger})~$ it is the system
with the selfinteraction described by the Hamiltonian
\begin{equation}\label{FCE2}
\widetilde {H}=\frac {1}{2}\,[\hat n\,f^{2}(\hat n)
+(\hat n+1)\,f^{2}(\hat n+1)]\,.
\end{equation}
Then the evolution operator
\begin{equation}\label{FCE3}
U(t)=\exp \,(-it\widetilde {H}(\hat n)\,)
\end{equation}
for this  quantum nonlinear oscillator gives the following solution 
to the Heisenberg equation of motion for the operator $~a\,(t)$
\begin{equation}\label{FCE4}
a\,(t)=U^{\dagger}(t)\,a\,U(t)=a\,\exp \,[-i\,\omega (\hat n)\,t]\,,
\end{equation}
where
\begin{equation}\label{FCE5}
\omega \,(\hat n)=\frac {1}{2}\,[(\hat n+1)\,f^{2}(\hat n+1)
-(\hat n-1)\,f^{2}(\hat n-1)]\,.
\end{equation}
Thus, we see then that also the quantum f--oscillator vibrates 
with a frequency depending on the amplitude.

Turning to the Schr\"odinger picture, we can remark that at time 
$~t~$ the harmonic oscillator has become a deformed oscillator of the 
kind we discuss. The deformation function is actually complex of 
modulus 1 and we will denote it with
\begin{equation}\label{FCE6}
F_{f}(n,\,t) =
\exp \left[- it~\frac { (n+1)\,f^{2}(n+1)
-(n-1)\,f^{2}(n-1)}{2}\right]\,.
\end{equation}
It is possible, in fact, to introduce the notion of f--coherent
states also for complex deformation functions $~f~$ as all formulae 
go through unaltered.

Then it can be seen, that if initially the state was the usual coherent 
state, i.e., in an eigenstate of the operator $~a\,,$ then it evolves
becoming at time $~t~$ an $~F_{f}({t})$--coherent state.  
Physically it means that f--nonlinearity creates the 
$~F_{f}({t})$--coherent states in the evolution of a usual coherent state.
Due to this, the photon statistics of the initial coherent state, to 
be discussed later, is influenced by the f--nonlinearity of the field 
vibrations. It is evidently different from the usual coherent states.     
Interesting physical example of the f--nonlinear systems is quartic
nonlinear oscillator usually used for modelling the Kerr medium.

\section{Irreducibility and Deformation}

\indent\indent
The usual Stone--von Neumann theorem states that the operators $~q~$ 
and $~p~$ (or $~a~$ and $~a^{\dagger}\,)~$ have no invariant subspaces 
in the Hilbert space of the oscillator states. If $~f\,(n)~$ is chosen 
to have no zeroes in $~\hbox {\bf Z}^{+},$ the operators $~A~$ and 
$~A^{\dagger}~$ are irreducible over the Fock space. If there are one 
or more double zeroes, the Fock space breaks up into a countable number 
of irreducible representations (compare with Master Analytic 
representations~\cite{sud2}\,). 
If the zeroes are simple zeroes, some of the reduced pieces will 
not allow a unitary resepresentation.

It is easy to prove, that if the function $~f\,(n)~$ has no zeroes at 
positive integers, the Stone--von Neumann theorem can be
extended to the case of the operators $~A,\,A^{\dagger}\,.$ So for the
q--oscillator, we are just in this case. Here the map $~a\rightarrow A~$
is invertible and the statement is obvious. More interesting situations
arise when, for example, the function $~f\,(n)~$ has one double zero 
at the integer $~n_{0}\,,$ i.e., $~f\,(n_{0})=f'(n_{0})=0\,.$ Then 
the subspace of the states
\begin{equation}\label{FCE7}
\mid \psi \rangle = \sum_{n=0}^{n_{0}} s_{n} \,\mid n\rangle
\end{equation}
is an invariant subspace for the operators $~A,\,A^{\dagger}~$ 
constructed by means of such function $~f\,(n)\,.$ The subspace of 
the states
\begin{equation}\label{FCE8}
\mid \psi'\rangle  = \sum_{n=n_{0}+1}^{\infty} s_{n} \mid n\rangle
\end{equation}
is another invariant subspace for the above operators. Thus, the
coherent states defined in this case do not contain the states with
photon numbers less or equal to $~n_{0}\,.$ In this case, the coherent 
state contains the states with photon numbers starting from 
$~n_{0}+1\,,$
\begin{equation}\label{FCE9}
\mid \alpha,\,\tilde f\rangle =\widetilde {N}\sum_{n=n_{0}+1}^{\infty}
\frac {\alpha^{n}}{\sqrt {\tilde {n}!}\,
[\tilde {f}(n)]!}\,\mid n\rangle \,,
\end{equation}
where
\begin{equation}\label{FCE10}
\tilde{n}! = n\,(n-1)\,(n-2)\cdots (n_{0}+1)\,,
\end{equation}
\begin{equation}\label{FCE11}
[\tilde{f}(n)]! = f(n)\,f(n-1)\cdots f(n_{0}+1)\,,
\end{equation}
and
\begin{equation}\label{FCE12}
\widetilde {N}_{\alpha, \,\tilde f}=\left(\sum_{n=n_{0}+1}^{\infty}
\frac {|\alpha|^{2n}}{\tilde {n}!\,
(\,[\tilde {f}(n)]!\,)^{2}}\right)^{-1/2}\,.
\end{equation}

\section{Completeness Relations}

\indent\indent
In this section, we will show how the f--coherent states form a 
complete system of states in the Hilbert space for nonsinguar 
$~f\,(n)\,,$ so that any state vector may be represented as 
a superposition of the f--coherent states. There may be different 
forms of completeness relations since the set of f--coherent states 
are over complete.

We introduce  first an integral representation for the identity
operator which uses the analyticity of the f--coherent states and 
the Cauchy theorem. By construction, the state
$~N_{f,\,\alpha}^{-1}\mid \alpha \rangle~$ is an analytic vector valued
function of the complex variable $~\alpha \,.$ Hence, the following 
relation holds
\begin{equation}\label{FCE13}
\mid n\rangle =\frac {\sqrt {n!}}{2\pi i}\,[f(n)]!\oint d\alpha ~
N^{-1}_{f,\,\alpha}
\,\mid \alpha, \,f\rangle \,\alpha ^{-n-1}\,.
\end{equation}
Inserting this formula into the known resolution of identity 
(\ref{AR5}) we obtain
\begin{equation}\label{FCE14}
\hbox {\bf 1}=-\frac {1}{4\pi ^{2}}
\sum _{n=0} ^{\infty} \oint \oint d\alpha ~d \beta^{*} 
\,\mid \alpha,\,f\rangle \langle \beta , \,f\mid \,N_{f,\,\alpha}^{-1}
\,N^{-1}_{f,\,\beta}\,(\alpha \beta ^{*})^{-n-1}n!\,(\,[f(n)]!\,)^2\,.
\end{equation}
These are line integrals along contours taken in a region of the 
complex planes where the convergence is guaranteed for the series 
considered in Section~7 to define the normalization constant 
$~N_{f,\,\alpha}\,.$ The introduced nondiagonal resolution of identity 
permits us to calculate the coefficients necessary to represent  
any vector as a superposition of f--coherent states.
We also point out that the  components of the f--coherent state
in any basis are the generating functions for components of the
Fock states. It means that for any matrix representing an operator
in Fock basis, the matrix elements of the same operator in f--coherent
state basis are the generating functions.

In order that such states can be considered as coherent states in the 
usual definition~\cite{ksmmb},
one should write a diagonal resolution of the identity
\begin{equation}\label{FCE15}
\mbox {\bf 1} = \int d\mu\,(\alpha)\,\mid \alpha,\,f\rangle \langle
\alpha,\,f\mid \,,
\end{equation}
where $~\mu\,(\alpha)~$ is the weight function.
Then the following relations have to be satisfied
\begin{equation}\label{FCE16}
2 \pi  \int_{0}^{\bar \rho} \rho^{2n+1}~ [N_{f}(\rho)\,]^{2} 
\mu\,(\rho)~d\rho = n!\,(\,[f(n)]!\,)^{2}\,;~~~~
\forall n\in \hbox {\bf Z}^{+}\,.
\end{equation}
These are actually an infinity of moment equations for the measure 
$~\mu \,.$ For the usual coherent states, as well as for the eigenstates 
of the q--deformed annihilation operators, such measures exist 
\cite{breg} and in both cases the integral is over the entire complex 
plane.

As far as the harmonious states of Section~7 are concerned, applying
the general resolution of identity (\ref{AR5}), one obtains
\begin{equation}\label{FCE17}
\hbox {\bf 1}=-\frac {1}{4 \pi^2}\oint\oint
\frac {d\alpha ~d\beta^{*}}{\alpha \beta^{*}-1} 
\frac {1}{\sqrt{(1-|\alpha|^{2})\,(1-|\beta|^{2})}}\mid\alpha, f_{h}
\rangle \langle \beta, f_{h}\mid 
\end{equation}
after having inserted the normalization constants (\ref{FQ22})
for the harmonious states $~\mid \alpha,\,f_{h}\rangle \,$ and 
$~\mid \beta,\,f_{h}\rangle \,.$
We can make this integral along two contours as an integral over 
the phases of $~\alpha ~$ and $~\beta \,,$ if the countours are 
chosen according to
\begin{equation}\label{FCE18}
\alpha=a\,e^{i\phi}\,;~~~~~~\beta=a\,e^{i\psi}\,,
\end{equation}
and $~0<a<1\,.$
Then the double contour integral transforms into
\begin{equation}
{\bf 1}=-\frac{a^2}{4\pi^{2}(1-a^2)^{2}}\int_{0}^{2\pi}
\int_{0}^{2\pi} d\phi ~d\psi
\frac {e^{i(\phi-\psi)}}{a^{2}\,e^{i(\phi-\psi)}-1}\,
\mid a\,e^{i\phi},\,f_{h}\rangle \langle a\,e^{i\psi},\,f_{h}\mid\,.
\end{equation}
We conclude remarking that for each $~a\in (0,\,1)~$ there is a 
completeness relation in  terms of projectors on harmonious states.

It is possible, however, to have a resolution of the identity
with the integral depending  on one parameter only, once states
with norm not necessarily  strictly positive are allowed. In~\cite{sud3}, 
such states have been considered after having introduced the following  
scalar product
$$
\langle \alpha \mid \beta \rangle =(1-\alpha ^{*}\beta)\,.
$$
Then, since
$$ 
\mid \alpha,\,f_{h}\rangle =(1-|\alpha |^{2})^{-1/2} \,
\sum_{n=0}^{\infty}\alpha ^{n}\,\mid n\rangle \,;~~~~
|\alpha|<1\,,
$$
we have
$$
\int \mid |\alpha |\,e^{i\theta}\rangle \langle |\alpha|
\,e^{i\theta}\mid
~\frac {d\theta}{2\pi}=\sum_{0}^{\infty} |\alpha|^{2n}\, 
(1-|\alpha|^{2})^{-1/2}\mid n\rangle \langle n\mid 
$$
and
$$
\int \mid |\alpha|^{-1}\,e^{i\theta}
\rangle \langle |\alpha|\,e^{i\theta}
\mid ~\frac{d\theta}{2\pi}
= (1-|\alpha|^{2})^{-1/2}(1-|\alpha|^{-2})^{-1/2}
\,\sum_{0}^{\infty} \mid n\rangle \langle n\mid \,.
$$
Hence,
$$
(2-|\alpha|^{2}-|\alpha|^{-2})^{1/2} 
\,\int \mid |\alpha|^{-1}\,e^{i\theta}\rangle \langle
|\alpha|\,e^{i\theta}\mid ~\frac{d\theta}{2\pi} =\hbox {\bf 1}\,.
$$

\section{Two-mode f-Coherent States}

\indent\indent
For usual multimode harmonic oscillators, there exist generalized 
correlated states~\cite{kur,bra} in which the mode quadratures are
statistically dependent. The quasidistributions for these states have
Gaussian form and the photon distribution function is described by
multivariable Hermite polynomials~\cite{olga}.
It is possible to extend nontrivially the construction of one-mode
f--oscillator to many modes. In particular, for two-mode state we 
consider the two operators
\begin{equation}\label{FS1}
A_{i} = a_{i}\,f\,(\hat n)\,;~~~~~i=1,\,2\,,
\end{equation}
where
\begin{equation}\label{FS2}
\hat n = \hat n_{1} + \hat n_{2}\,;
~~~~~\hat n_{i} = a_{i}^{\dagger}\,a_{i}
\end{equation}
(the operators $~a_i~$ satisfy boson commutation relation).
These operators commute and for this reason we can construct 
algebraically the two-mode f--coherent state 
$~\mid \alpha_{1},\,\alpha_{2},\,f\rangle ~$ defined by the 
following equations
\begin{equation}\label{FS3}
A_{i}\mid \alpha_{1},\,\alpha_{2},\,f\rangle =
\alpha_{i}\mid \alpha_{1},\,\alpha_{2},\,f\rangle \,,
~~~~~~i=1,\,2\,.
\end{equation}
Considering the series expansion
\begin{equation}\label{FS4}
\mid \alpha_{1},\,\alpha_{2},\,f\rangle =\sum_{n_{1}=0}^{\infty}\,
\sum_{n_{2}=0}^{\infty}\,c_{n_{1},\,n_{2}}~\mid n_{1},\,n_{2}\rangle \,,
\end{equation}
where the Fock states $~\mid n_{1},\,n_{2}\rangle ~$ satisfy
\begin{equation}\label{FS5}
a_{1}^{\dagger}\,a_{1}\mid n_{1}\,,n_{2}\rangle 
= n_{1}\mid n_{1},\,n_{2}\rangle \,;
~~~~~n_{1} \in \hbox {\bf Z}^{+}
\end{equation}
and
\begin{equation}\label{FS6}
a_{2}^{\dagger}\,a_{2}\mid n_{1},n_{2}\rangle
= n_{2}\mid n_{1},\,n_{2}\rangle \,;~~~~~n_{2}\in \hbox {\bf Z}^{+}\,.
\end{equation}
The solution of the recurrence relation which is obtained for the
$~c_{n_{1}n_{2}}$'s is
\begin{equation}\label{FS7}
c_{n_{1}n_{2}} = c_{00}\,\frac {\alpha_{1}^{n_{1}}
 \alpha_{2}^{n_{2}}}{\sqrt {n_{1}!n_{2}!}\,[f(n)]!}
\end{equation}
with $c_{00}\,,$ fixed as before by  normalization, being
\begin{equation}\label{FS8}
c_{00} =\left (\sum_{n_{1}=0}^{\infty} \sum_{n_{2}=0}^{\infty} 
|\alpha_{1}|^{2n_{1}}\,
|\alpha_{2}|^{2n_{2}}\,(\,[f(n)]!\,)^{-2}\,(n_{1}!\,n_{2}!)^{-1}
\right )^{-1/2}\,.
\end{equation}
The two-mode f--coherent state can be now defined as
\begin{equation}\label{FS9}
\mid \alpha_{1},\,\alpha_{2},\,f\rangle = c_{00}\,\sum_{n_{1}=0}^{\infty}
\sum_{n_{2}=0}^{\infty}
\frac{\alpha_{1}^{n_{1}}\alpha_{2}^{n_{2}}}
{\sqrt{n_{1}!\,n_{2}!}\,[f(n)]!}
\,\mid n_{1},\,n_{2}\rangle \,;~~~~~\alpha_{i}\in \hbox {\bf C}\,;
~~~i=1,\,2\,.
\end{equation}
It should be remarked that in this form there is a coupling between 
the two modes, as there is a dependence of each of them on the total
energy, this interaction between the two modes in general is 
nonlinear.

Another generalization for two-mode coherent states is of course 
obtained by means of the product of two one-mode f--coherent states, 
so that there is no interaction between the two modes. After defining
\begin{equation}\label{FS10}
A_{i}^{'} = a_{i} f_{i}(n_{i})\,;~~~~i= 1,\,2
\end{equation}
and finding their eigenstates, we can make
the tensor product obtaining
\begin{eqnarray}\label{FS11}
\mid\alpha_{1}, \alpha_{2}, f_{1}, f_{2}\rangle
=\left(\sum_{n_{1}=0}^{\infty} \sum_{n_{2}=0}^{\infty}
|\alpha_{1}|^{2n_{1}}\,|\alpha_{2}|^{2n_{2}}\,
(n_{1}!\,n_{2}!)^{-1}\,
(\,[f_{1}(n_{1})]!\,[f_{2}(n_{2})]!\,)^{-2}\right)^{-1/2}\nonumber\\
\otimes \sum_{n_{1}=0}^{\infty} \sum_{n_{2}=0}^{\infty}
\frac {\alpha_{1}^{n_{1}} \alpha_{2}^{n_{2}}}
{\sqrt{n_{1}!\,n_{2}!}\,[f_{1}(n_{1})]!\,[f_{2}(n_{2})]!}\,
\mid n_{1},\,n_{2}\rangle \,.
\end{eqnarray}
In the case $~f_{1} =  f_{2} = 1\,,$
we have the usual two-mode coherent states, namely,
\begin{equation}\label{FS13}
\mid \alpha_{1},\,\alpha_{2},\,1,\,1\rangle
= e^{-(|\alpha_1|^2 + |\alpha_2|^2)/2}\,
\sum_{n_{1}=0}^{\infty} \sum_{n_{2}=0}^{\infty} \frac {\alpha_1^{n_1}
\alpha_2^{n_2}}{\sqrt {n_{1}!\,n_{2}!}}\,\mid n_{1},\,n_{2}\rangle \,.
\end{equation}

\section{Physical Application of f-coherent States}

\indent\indent
We give the examples of how the notions so far discussed might be of 
some interest in dealing with physical problems. With the choice 
made and repeatedly illustrated, we can continue interpreting
$~\mid n\rangle ~$ as the state containing $~n~$ quanta, also when 
it appears in a 
series representing an f--coherent state. Recalling the wide use 
made in quantum optics of the harmonic oscillator formalism,   
denoting by $~\mid n\rangle ~$ a state containing $~n$ photons,   
our examples will all be related to this interpretation.

In the usual coherent state $~\mid \alpha,\,1\rangle \,,$ the particle 
distribution is given by the Poissonian function
\begin{equation}\label{PD1}
P_{1,\,\alpha}(n) = \frac {|\alpha|^{2n}}{n!}\,e^{-|\alpha|^{2}}
\end{equation}
with mean photon number
\begin{equation}\label{PD2}
\langle n\rangle _{1} = |\alpha|^{2}
\end{equation}
and dispersion
\begin{equation}\label{PD3}
\sigma_{1,\,n} = \langle n^{2}\rangle _{1} 
-\langle n\rangle _{1}^{2}  = |\alpha|^{2}\,.
\end{equation}
Then the ratio of the above quantities 
$~\sigma_{1,\,n}/\langle n\rangle _{1} = 1\,.$

In the case of f--coherent state the above equations become
\begin{equation}\label{PD5}
P_{f,\,\alpha}(n) = |c_{n}|^{2} = \left(\sum_{j=0}^{\infty} 
\frac {|\alpha|^{2j}}{j!\,(\,[f(j)]!\,)^2}\right)^{-1}
\frac {|\alpha|^{2n}}{n!\,(\,[f(n)]!\,)^{2}}\,.
\end{equation}
The mean photon number and dispersion are
\begin{equation}\label{PD6}
\langle n\rangle _{f} = \sum_{n=0}^{\infty} n\,P_{f,\,\alpha}(n)
\end{equation}
and
\begin{equation}\label{PD7}
\sigma_{f,\,n} = \sum_{n=0}^{\infty} n^{2}P_{f,\,\alpha}(n)  
-\left(\sum_{n=0}^{\infty}n\,P_{f,\,\alpha}(n)\right)^{2}\,,
\end{equation}
while their ratio $~\sigma_{f,\,n}/\,\langle n\rangle _{f}~$
can be less or greater than 1, producing for a given $~f~$ either 
sub-Poissonian or super-Poissonian statistics.

For the case considered in Section~10, where $~f~$ has zeroes, 
the photon distribution function $~P\,(n)~$ vanishes for 
$~n\leq n_{0}$
and
\begin{equation}\label{PD9}
P\,(n) = \frac {\tilde {N}^{2}\,|\alpha|^{2n}}{\tilde{n}!\,
(\,[\tilde{f}(n)]!\,)^{2}}\,;~~~ n>n_{0}\,.
\end{equation}
The photon distribution in the first of the two generalization for 
two modes (\ref{FS1}) is
\begin{equation}\label{PD10}
P_{f,\,\alpha_{1},\,\alpha_{2}}(n_{1},\,n_{2})= c_{00}^{2}\,
 \frac {|\alpha_1|^{2n_1}\,|\alpha_2|^{2n_2}}{n_1!\,n_2!\,
(\,[f(n)]!\,)^{2}}
\end{equation}
and for the second one (\ref{FS10}) reads
\begin{equation}\label{PD11}
P_{f_{1},\,f_{2},\,\alpha_{1},\,\alpha_{2}}
=P_{f_{1},\,\alpha_{1}}\,P_{f_{2},\,\alpha_{2}}\,.
\end{equation}
We emphasize the fact that in the latter case the two modes are
independent and there is no correlation.

On the contrary, in the previous two-mode generalization a correlation 
exists between the modes: in facts if one gives the interpretation to 
the f--coherent states as the states related to  deformation of the 
annihilation operator due to a specific f--nonlinearity, in the first 
case, this happens through the total energy and then we can now 
conclude that this particular nonlinearity of the field produces 
a correlation among the modes.
In the case of q--oscillators, we obtain
for the photon distribution in the one-mode q--coherent state 
\begin{equation}\label{PD13}
P_{f_{q},\,\alpha}(n) =\left(\sum_{j=0}^{\infty}
\,\frac {|\alpha|^{2j}}{[
\frac{\sinh \,\lambda j}{\sinh \,\lambda}]!}\right)^{-1} 
\frac{|\alpha|^{2n}}{[\frac {\sinh \,\lambda n}
{\sinh \,\lambda}]!}\,.
\end{equation}
The property of this distribution is that, for large 
$~n~(n\gg \lambda^{-1})\,,$ the probability to have $~n~$ photons 
differs essentially from Poissonian distribution due the exponentially
decreasing of the denominator. The mean photon number 
$~\langle n\rangle _{q}~$ is given for q--nonlinear field by
\begin{equation}\label{PD14}
\langle n\rangle _{q}=\left(\sum_{j=0}^{\infty}\,\frac {|\alpha|^{2j}}
{[\frac {\sinh \,\lambda j}{\sinh \,\lambda}]!}\right)^{-1}
\sum _{n=0}^{\infty }\,\frac{n\,|\alpha|^{2n}}{[
\frac {\sinh \,\lambda n}{\sinh \,\lambda}]!}\,,
\end{equation}
the second moment by
\begin{equation}\label{PD15}
\langle n^{2}\rangle _{q} =\left(\sum_{j=0}^{\infty}\,
\frac {|\alpha|^{2j}}{[\frac
{\sinh \,\lambda j}{\sinh \,\lambda }]!}\right)^{-1}
\sum _{n=0}^{\infty }              
\,\frac{n^2\,|\alpha|^{2n}}{[\frac {\sinh \,\lambda n}
{\sinh \,\lambda }]!}\,,
\end{equation}
and the dispersion depends of course on the nonlinearity parameter 
$~\lambda \,.$

The photon distribution demonstrates the fast decreasing of 
the distribution function in comparison with the Poisson distribution 
function; the q--nonlinearity makes photon statistics sub-Poissonian. 
The q--nonlinearity and some other types of f--nonlinearities 
may influence the statistical properties that can be checked by 
Hanbury Brown--Twiss-like experiments and the presumed
detection of sub-Poissonian counting distribution in quantum optics.

\section{Squeezing and Correlation}
\indent\indent
Now we will calculate the squeezing and correlation of the quadrature
components in the introduced f--coherent states. We face, in fact, 
the problem since the discussed nonlinearity (for example, 
q--nonlinearity) of the field produces the state which is 
f--coherent state. Then this nonlinearity yields the phenomenon of 
squeezing and correlation of the field (photon) quadrature 
components. It is possible to calculate the dispersion and correlation 
of the quadratures explicitly. To do this we will take advantage of 
Eq.~(\ref{AR9}), which gives the expression
\begin{eqnarray}\label{scf1}
a&=&\frac {1}{f\,(\hat{n}+1)}\,A\,;\nonumber\\
\\
a^{\dagger} &=&A^{\dagger}\,\frac {1}{f\,(\hat{n}+1)}\,.\nonumber
\end{eqnarray}
Then the quadrature mean values are expressed through
\begin{equation}\label{scf2}
\langle \alpha ,\,f\mid a\mid \alpha ,\,f\rangle
=\alpha \,N_{\alpha ,\,f}^{2}\sum _{n=0}^{\infty }
\frac {|\alpha |^{2n}}{f(n+1)\,n!\,\{\,[f(n)]!\,\}^{2}}\,,
\end{equation}
yelding
\begin{equation}\label{scf3}
\langle \alpha ,\,f\mid x\mid \alpha ,\,f\rangle
=\frac {\alpha +\alpha ^{*}}{\sqrt 2}\,
N_{\alpha ,\,f}^{2}\sum _{n=0}^{\infty }
\frac {|\alpha |^{2n}}{f(n+1)\,n!\,\{\,[f(n)]!\,\}^{2}}\,,
\end{equation}
\begin{equation}\label{scf4}
\langle \alpha ,\,f\mid p\mid \alpha ,\,f\rangle
=\frac {\alpha -\alpha ^{*}}{i\sqrt 2}\,
N_{\alpha ,\,f}^{2}\sum _{n=0}^{\infty }
\frac {|\alpha |^{2n}}{f(n+1)\,n!\,\{\,[f(n)]!\,\}^{2}}\,.
\end{equation}
The dispersions may be calculated from the relations
\begin{equation}\label{scf5}
\langle \alpha ,\,f\mid a^2\mid \alpha ,\,f\rangle
=\alpha ^{2}N_{\alpha ,f}^{2}\sum _{n=
=0}^{\infty }
\frac {|\alpha |^{2n}}{f(n+1)\,f(n+2)n!\,\{\,[f(n)]!\,\}^{2}}\,,
\end{equation}
\begin{equation}\label{scf6}
\langle \alpha ,\,f\mid a^{\dagger}a\mid \alpha ,\,f\rangle 
=|\alpha |^{2}N_{\alpha ,\,f}^{2}
\sum _{n=0}^{\infty }
\frac {|\alpha |^{2n}}{f^{2}(n+1)\,n!\,\{\,[f(n)]!\,\}^{2}}\,,
\end{equation}
as
\begin{eqnarray}\label{scf7}
\langle \alpha ,\,f\mid x^2\mid \alpha ,\,f\rangle
&=&\left.\frac {1}{2}\left \{1+N_{\alpha ,\,f}^{2}\left [2\,|\alpha |^{2}
\right.\right.\right.
\sum _{n=0}^{\infty }
\frac {|\alpha |^{2n}}{f^{2}(n+1)\,n!\,\{\,[f(n)]!\,\}^{2}}\nonumber\\
&+&(\alpha ^{2}+\alpha ^{*2})\sum _{n=0}^{\infty }\left.\left.
\frac {|\alpha |^{2n}}{f(n+1)\,f(n+2)\,n!\,\{\,[f(n)]!\,\}^{2}}\right]
\right \}\,,
\end{eqnarray}
\begin{eqnarray}\label{scf8}
\langle \alpha ,\,f\mid p^2\mid \alpha ,\,f\rangle
&=&\left.\frac {1}{2}\left \{1+N_{\alpha ,\,f}^{2}\left [2\,|\alpha |^{2}
\right.\right.\right.
\sum _{n=0}^{\infty }
\frac {|\alpha |^{2n}}{f^{2}(n+1)\,n!\,\{\,[f(n)]!\,\}^{2}}\nonumber\\
&-&(\alpha ^{2}+\alpha ^{*2})\sum _{n=0}^{\infty }\left.\left.
\frac {|\alpha |^{2n}}{f(n+1)\,f(n+2)\,n!\,\{\,[f(n)]!\,\}^{2}}\right]
\right \}\,.
\end{eqnarray}
Thus, for quadrature dispersion
$~\sigma _{x}=\langle \alpha ,\,f\mid x^2\mid \alpha ,\,f\rangle
-\langle \alpha ,\,f\mid x\mid \alpha ,\,f\rangle ^2,$
we have
\begin{equation}\label{scf9}
\sigma _{x}=\frac {1}{2}+\mu _{x}\alpha ^{2}+\mu _{x}^{*}\alpha ^{*2}+
\nu _{x}\alpha \alpha ^{*},
\end{equation}
where
\begin{equation}\label{scf10}
\mu _{x}=\frac {1}{2}\,
N_{\alpha ,\,f}^{2}
\left \{\sum _{n=0}^{\infty }
\frac {|\alpha |^{2n}}
{f(n+1)f(n+2)n!\{[f(n)]!\}^{2}}
-N_{\alpha ,\,f}^{2}
\left (\sum _{n=0}^{\infty }
\frac {|\alpha |^{2n}}
{f(n+1)n!\{[f(n)]!\}^{2}}\right )^{2}\right \}
\end{equation}
and
\begin{equation}\label{scf11}
\nu _{x}=N_{\alpha ,\,f}^{2}\left \{\sum _{n=0}^{\infty }
\frac {|\alpha |^{2n}}{f^{2}(n+1)n!\{[f(n)]!\}^{2}}
-N_{\alpha ,\,f}^{2}\left (\sum _{n=0}^{\infty }
\frac {|\alpha |^{2n}}{f(n+1)n!\{[f(n)]!\}^{2}}\right )^{2}\right \}\,.
\end{equation}
For the other quadrature dispersion
$~\sigma _{p}
=\langle \alpha ,\,f\mid p^2\mid \alpha ,\,f\rangle
-\langle \alpha ,\,f\mid p\mid \alpha ,\,f\rangle ^2,$
we have
\begin{equation}\label{scf12}
\sigma _{p}=\frac {1}{2}+\mu _{p}\alpha ^{2}+\mu _{p}^{*}\alpha ^{*
2}+\nu _{p}\alpha \alpha ^{*},
\end{equation}
where
\begin{equation}\label{scf13}
\mu _{p}=-\frac {1}{2}\,N_{\alpha ,\,f}^{2}\left \{\sum _{n=0}^{\infty }
\frac {|\alpha |^{2n}}{f(n+1)f(n+2)n!\{[f(n)]!\}^{2}}
-N_{\alpha ,\,f}^{2}\left (\sum _{n=0}^{\infty }
\frac {|\alpha |^{2n}}{f(n+1)n!\{[f(n)]!\}^{2}}\right )^{2}\right \}
\end{equation}
and
\begin{equation}\label{scf14}
\nu _{p}=N_{\alpha ,\,f}^{2}\left \{\sum _{n=0}^{\infty }
\frac {|\alpha |^{2n}}{f^{2}(n+1)n!\{[f(n)]!\}^{2}}
-N_{\alpha ,\,f}^{2}\left (\sum _{n=0}^{\infty }
\frac {|\alpha |^{2n}}{f(n+1)n!\{[f(n)]!\}^{2}}\right )^{2}\right \}\,.
\end{equation}
Depending on the function $~f\,(n)$ the dispersion $~\sigma _{x}$
($~\sigma _{p}$) may become less than 1/2\,. It means
squeezing. One can calculate correlation of the quadrature 
components in f--coherent states as
\begin{equation}\label{scf15}
\sigma _{xp}=\frac {\alpha ^{2}
-\alpha ^{*2}}{2i}\,N_{\alpha ,\,f}^{2}\left [\sum _{n=0}^{\infty }
\frac {|\alpha |^{2n}}{f(n+1)f(n+2)n!\{[f(n)]!\}^{2}}
-N_{\alpha ,\,f}^{2}\left (\sum _{n=0}^{\infty }
\frac {|\alpha |^{2n}}{f(n
+1)n!\{[f(n)]!\}^{2}}\right )^{2}\right ].
\end{equation}
Then the quadrature correlation coefficient
$~r=\sigma _{xp}/\sqrt {\sigma _{x}\,\sigma _{p}}~$
is not equal to zero. Thus, the f--coherent state has the property
of being a correlated state \cite{kur}. The invariant
$\sigma _{x}\sigma_{p}-\sigma _{xp}^{2}$
is larger than 1/4. Thus, the f--coherent states {\it do not} 
minimize the Schr\"odinger uncertainty relation
\cite{schrod,robert}.

\section{Deformation of Planck Formula}
\indent\indent
We will discuss what physical consequences may be found if the 
considered f--nonlinearity influences the vibrations of the real 
field mode oscillators like, for example, electromagnetic field 
oscillators or the oscillations of the nuclei in polyatomic molecules.

First, it will be seen that this nonlinearity changes the specific 
heat behaviour. To show this, we have to find the partition function 
for a single f--oscillator corresponding to the Hamiltonian 
$~H=(AA^{\dagger} + A^{\dagger}A)/2~$
\begin{equation}\label{1dp1}
Z(T)=\sum_{n=0}^{n=\infty} \exp (-\beta E_{n})\,,
\end{equation}
where the variable $~\beta $
is the inverse temperature $~T^{-1}$ and $E_{n}$ was
given in Eq.~(\ref{NO2}). For the evaluation of the quantum
partition function for an ensemble of q--oscillators~\cite{sol1}, 
we first note that in this case
$$
E_{n}=\frac {1}{\sinh \,\lambda}\,[\sinh \,\lambda (n+1) 
+ \sinh \,\lambda n]\,;
~~~\lambda = \log q\,,
$$
and obtain that the specific heat decreases for 
$~T\rightarrow \infty $ as
\begin{equation}\label{1dp2}
C\propto \frac{1}{\ln~T}\,.
\end{equation}
Thus, the behaviour of the specific heat of the q--oscillator
is different from the behaviour of the usual oscillator 
in the high temperature limit. This property may
serve as an experimental check of the existence of vibrational
nonlinearity of the q--oscillator fields.

The q--deformed Bose distribution can be obtained by the same
method and one has~\cite{sol1}
\begin{equation}\label{1dp3}
\langle n\rangle =\bar n_{0}-\beta \,\frac{\lambda ^{2}}{6}
\left[\,\frac{1}{2}\,
\left ((\overline {n^2})_0-(\bar n)^{2}_0\right )
+\frac{3}{2}\,\left ((\overline {n^3})_0
=-\bar {n}_{0}(\overline {n^2})_0\right )
+(\overline {n^4})_0-\bar {n}_{0}(\overline {n^3})_0\right]\,,
\end{equation}
in which $~\bar{n}_0$ is the usual Bose distribution function and
\begin{equation}\label{1dp4}
(\overline {n^k})_0 = 2\,\sinh \,\frac{\beta}{2} \sum_{n=0}^{\infty}
 n^{k}\,e^{-\beta (n+1/2)}\,.
\end{equation}

Calculating the partition function for small q--nonlinearity parameter
we have also the following q--deformed Planck distribution formula
\begin{equation}\label{1dp5}
\langle n\rangle =\frac {1}{e^{\hbar \omega /kT}-1}-\lambda ^{2}\,
\frac {\hbar \omega}{kT}\,\frac {e^{3\hbar \omega /kT}+4e^{2\hbar \omega /kT}
+e^{\hbar \omega /kT}}{(e^{\hbar \omega /kT}-1)^{4}}\,.
\end{equation}
It means that q--nonlinearity deforms the formula for the mean photon
numbers in black body radiation~\cite{sol1}. 

One can write down the high and low temperature approximations for
the deformed Planck distribution formula \cite{sol3}.           
For small temperature, the behaviour of the deformed Planck 
distribution differs from the usual one
\begin{equation}\label{1dp6}
\langle n\rangle -\bar n_{0} = -\lambda ^{2}\,\frac {\hbar \omega }{kT}\,
e^{-\hbar \omega /kT}\,.
\end{equation}
For the high temperature, the nonlinear correction to the usual mean 
photon number also depends on temperature
\begin{equation}\label{1dp7}
\langle n\rangle -\bar n_{0}  = - \lambda ^{2}\,\left (
\frac {\hbar \omega }{kT}\right )^{-3}\,.
\end{equation}
As it was seen, the discussed q--nonlinearity 
produces a correction to the Planck distribution formula (mean oscillator 
energy) and this may also be subjected to an experimental test.

As it was suggested in~\cite{sol2}, the q--nonlinearity of the field
vibrations produces blue shift effect which is the effect of the
frequency increase with the field intensity. For small nonlinearity
parameter $~\lambda $ and for large number of photons $~n~$ in
a given mode, the relative shift of the light frequency is
$$
\frac {\delta \omega }{\omega }
=\frac {\lambda ^{2}\,n^{2}}{2}\,.
$$
This consequence of the possible existence of a q--nonlinearity may
be relevant for models of the early stage of the Universe.

Another possible phenomenon related to the q--nonlinearity was
considered in~\cite{marman}, where it was shown, that if one deforms
the electrostatics equation using the method of deformed creation
and annihilation operators, a point charge acquires a formfactor
due to q--nonlinearity.

\section{Conclusion}
\indent\indent
Starting with the example of the harmonic oscillator, we have
exhibited a family of associated nonlinear systems which are
completely integrable, both in classical and quantum physics.

We have shown that q--nonlinearity, associated with quantum 
groups, is a subclass of a more general class of possible 
nonlinearity. These aspects, related to the existence of alternative 
Hamiltonian descriptions for the harmonic oscillator, have been 
considered with respect to the consequence for the partition function, 
again both in the classical and quantum situation. We have found
that while  the partition function for the harmonic oscillator
does not depend on the particular Hamiltonian, for the nonlinear
ones it does depend, giving therefore an experimental possibility
to select among them.

A class of states has been considered in the Fock space through
the deformation process applied to the harmonic oscillator operators.
Such states have been described as f--coherent states (or nonlinear
coherent states), harmonious states and q--coherent states being 
particular examples of them. Their
different representation have been constructed, like the Wigner--Moyal
and Husimi--Kano distributions. It is shown how nonlinear couplings 
between different modes are easy to obtain.

Keeping unaltered the current physical identification of the Fock
states, a possible use is presented in the field of quantum optics, 
obtaining deformed photon distributions and related physical quantities.
Physical consequences of the deformed vibrations,
like the Planck distribution deformation, are then reviewed for
the q--oscillators, where a blue shift effect exists.
The related phenomena were studied recently~[36--38].
%\cite{ex1,ex2,ex3}.
In~\cite{ex1}, the estimation of upper limit of q--nonlinearity of the
electromagnetic field vibration was done.

The studied nonlinearities, if they exist, for the electromagnetic
field or for the gluons, may influence the particle decays,
correlations in particle multiplicities, and a change in the 
Hanbury Brown--Twiss experiment results.
It would naturally affect the stimulated emission
rates and hence the radiative equilibrium in the presence of matter.

\section*{Acknowledgments}
\indent\indent
One of us (V. I. M.) would like to acknowledge Universit\'a di Napoli
``Federico II'' and Osservatorio Astronomico di Capodimonte
for their kind hospitality.

\pagebreak

\end{document}